\documentclass[english,prl,twocolumn,showpacs,floatfix]{revtex4}
\usepackage[english]{babel}
\usepackage{epsfig}
\usepackage{amsmath}
\usepackage{amssymb}
\usepackage{amsfonts}
\usepackage{array}
\usepackage{subfigure}
\usepackage{psfrag}

\DeclareMathOperator{\e}{e}
\DeclareMathOperator{\erf}{erf}

\begin{document}
\title{Scaling in rupture of polymer chains}
\author{ S. Fugmann and I.M. Sokolov}

\affiliation{Institut f\"{u}r Physik, Humboldt-Universit\"{a}t
  Berlin, Newtonstrasse 15, 12489 Berlin, Germany}

\begin{abstract}
We consider the rupture dynamics of a homopolymer chain pulled at one
end at a constant loading rate $r$.  Compared to single bond breaking,
the existence of the chain introduces two new aspects into rupture
dynamics: the non-Markovian aspect in the barrier crossing and the
slow-down of the force propagation to the breakable bond. The relative
impact of both these processes is investigated, and the second one was
found to be the most important at moderate loading rates.  The most
probable rupture force is found to decrease with the number of bonds
as $f_{max}\propto const-(\ln (N/r))^{2/3}$ and finally to approach a
saturation value independent on $N$. All our analytical findings are
confirmed by extensive numerical simulations.
\end{abstract}

\maketitle

Modern developments of micromanipulation techniques made possible the
experiments on the mechanical response of single molecules to a
well-defined load. The typical setup here is the load ramping in which
the force grows linearly in time until the molecule breaks or changes
its structure.  Examples are single molecule pulling experiments
\cite{Rief97,Mehta99,Grandbois99,Cui06,Friedsam03,Neuert07} and the
ones on protein unfolding \cite{Strunz99}. The dynamic force
spectroscopy \cite{Evans01,Strunz99} delivering the complete spectrum
of the rupture force versus the loading rate offers an extremely
powerful tool to determine the bonds' strengths and gives deep
insights into the internal dynamics of the structure under study
\cite{Garg95,Dudko03,Dudko06,Chen05,Dias05}.

From the theoretical point of view the rupture of a bond under a
constant loading rate can be described as a thermally activated escape
process from a time-dependent potential well
\cite{Evans97,Chen05,Dudko03,Dudko06,Lin07,Friddle08}. Dudko et al. \cite{Dudko03}
approximated the energy landscape close to the barrier up to the third
order and predicted that the rupture force scales like $const+(\ln
r)^{2/3}$, with $r$ being the loading rate. It is shown
\cite{Dudko03}, that particularly in the strong pulling limit
\cite{Lin07,Friddle08} this theory reveals better agreement with
numerical simulations then a linear theory given in
\cite{Evans97,Evans01}.

In application to polymers this theoretical description
concentrates on breaking of \textit{one} bond (presumably the weakest
one) and fully disregards the chain structure of the system. On the
other hand, the situation of breaking of the more or less homogeneous
chain has hardly been considered. Thus, Ref.~\cite{Dias05} discusses
breaking of a ring of identical bonds, i.e., of a chain with periodic
boundary conditions. We note that although the experimental
realization of this situation is possible in the setup discussed in
Ref.~\cite{Severin06}, the typical case corresponds to a linear chain
pulled at one of its ends with another end clamped. In what follows we
consider this typical situation where the force is applied to one of
the terminals of the chain while the other end is grafted to a
surface.

For sufficiently low loading rates and short chains each bond
experiences the same force and the rupture can occur at an arbitrary
bond. For a longer chain or for a high loading rate the actual force
profile along the chain has to be taken into account. When pulling at
one terminal of the chain the time of the order of the Rouse time
$\tau_R$ \cite{DoiEdwards86} is necessary for the stress to propagate
through the chain; hence the stress in adjacent bonds differs. If bond
rupture occurs on a time scale shorter than the Rouse time, only a
part of the chain is under stress and accounts for the rupture
process. Thus the rupture force will be crucially affected by the
number of monomers in the chain. The aim of this work is to
investigate the impact of the chain length on the rupture force both
numerically and analytically. Our model will correspond to a chain of
monomers interacting via the Morse
potential. Otherwise, the model is identical to the Rouse one: we
disregard hydrodynamical interactions and describe the interaction of
the monomers with the heat bath via independent white noises.

The monomer-monomer interaction along the chain is modellled by 
the Morse potential $U(q)$ given by
\begin{equation} U(q)=\frac{C}{2\alpha} \left(1-e^{-\alpha \left(
q\right)}\right)^2\,,
\label{eq:morse_pot}
\end{equation} with dissociation energy $C/(2\alpha)$ and stiffness
$C \alpha$. The constant loading enters through an additional time dependent
potential of the form $L(q,t)=-qRt$ with loading rate $R$. We note that this
potential differs from the often used harmonic linker potential
\cite{Evans97,Dudko03,Chen05,Tshiprut06}, but preserves the feature of
linearly increasing force within time we are focusing on. Its
advantage is that the extrema of the overall potential
$U(q,t)=U(q)+L(q,t)$ can be obtained analytically.
To illustrate the viability of the effects let us take
the experimentally relevant values of parameters --
$C/(2\alpha)\sim 10^{-19} J$ with $\alpha\sim 10^{10}m^{-1}$, the friction
$\gamma=10^{-6}kg/s$, and a loading rate $R=10^{-7}N/s$ and obtain, that 
the effects discussed below are important for chains 
consisting of around 300 monomers, i.e. already for relatively short ones.

To proceed, let us first recall the rupture dynamics of a single Morse bond
under constant loading
\cite{Dudko03,Raible04,Chen05,Dudko06,Tshiprut06}. 
At smaller loads the overall potential has two extrema, a minimum
corresponding to a metastable state of the pulled bond, and a maximum
providing the activation barrier. There exists a critical load
$F_i=F(t_i)=C/4$ for which the extrema merge at $q_i=\ln(2)/\alpha$ and disappear. In the
purely deterministic dynamics the Morse bond breaks exactly at
$t_i=F_i/R$. Since the system is in contact to a heat bath at temperature $T$, its overdamped dynamics is described by
$\dot{q}=-C/\gamma(1-\exp(-\alpha q))\exp(-\alpha q)+R/\gamma t+\sqrt{2k_BT/\gamma}\xi(t)$, with $\xi(t)$ being Gaussian white noise. We introduce $c=C/\gamma$ ($\left[c\right]=nm/\mu s$), $r=R/\gamma$ ($\left[r\right]=nm/\mu s^2$) and $f=F/\gamma$ (in the following $f$ is referred to as force). The diffusion coefficient is denoted by $D=k_BT/\gamma$ ($\left[D\right]=nm^2/\mu s$). Then, the above given experimentally relevant loading rate corresponds to $r=10^{-4}nm/\mu s^2$ in our calculations.

The probability $W_1(t)$ that the bond remains intact
can be expressed through the following kinetic equation
\cite{Tshiprut06,Dudko03,Dudko06,Raible04} ${dW_1(t)}/{dt}=-k(t)W_1(t),$
with $k(t)$ being the Kramers rate \cite{Kramers40,Hanggi90}. 
Taking $f=rt$ we can rewrite the kinetic equation in the form
\begin{equation}
\frac{dW_1(f)}{df}=-\frac{1}{r}k(f)W_1(f)\,.
\label{eq:kinetic3}
\end{equation}
The measured PDF for the rupture forces $P_1(f)$ then is $P_1(f)=-dW_1(f)/df$.

Under the assumption that $f$ is close to $f_i$ when bond rupture
occurs, it is usual to expand the potential around the inflection
point $q_i$ up to the third order in deviations from $q_i$
\cite{Garg95,Dudko03,Dudko06}. The Kramers rate becomes $k(f)=c\alpha/(4\pi)\sqrt{1-f/f_i}\exp\left(-c/(3\alpha D)(1-f/f_i)^{3/2}\right)$. Solving \eqref{eq:kinetic3} one derives
\begin{equation}
W_1(f)=\exp\left\{-\frac{v}{r}\exp\left[-w\left(1-\frac{f}{f_i}\right)^{\frac{3}{2}}\right]\right\}
\label{eq:w1}
\end{equation}
with $v=c\alpha^2D/(8\pi)$ and $w=c/(3\alpha D)$. 
In the limit of small loading rates the most probable rupture force
$f_{max}$ follows the scaling relation \cite{Dudko03}
\begin{equation}
f_{max}=f_i\left[1-\left(\frac{\ln\left(v/r\right)}{w}\right)^{\frac{2}{3}}\right]\,.
\label{eq:scaling}
\end{equation}
%

\begin{figure}
\begin{center}
\includegraphics[height=5.5cm,width=7.cm]{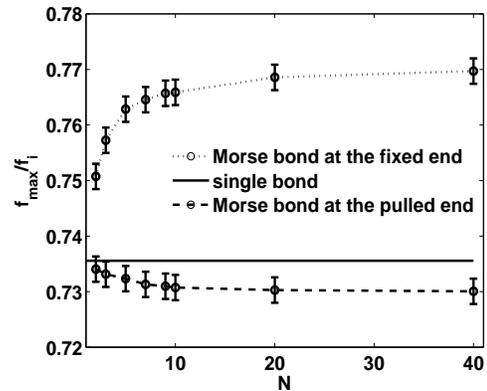}
\end{center}
\caption{\label{fig:mean_n} Most probable rupture force $f_{max}$ as a
function of the chain length $N$. The chain consist of $N-1$ harmonic
springs and one breakable Morse bond, either situated at the fixed
wall or at the pulled terminal of the chain. The coupling parameter of
the harmonic springs $\tilde{c}$ is obtained via the first order
series expansion of the Morse potential given in
Eq.~\ref{eq:morse_pot}, i.e., $\tilde{c}=\alpha c$. Superimposed is
the value of $f_{max}$ obtained from a single bond pulling
experiment. The loading rate is $r=10^{-5}nm/\mu s^2$. The remaining parameter values are $c=3.5nm/\mu s$,
$D=2\times10^{-3}nm^2/\mu s$, and $\alpha=10/nm$. Error bars indicate the uncertainty due to the binning of numerical data.}
\end{figure}

Let us now turn to a chain of $N$ bonds. One of its ends is fixed and
another one is exposed to a monotonously increasing force with loading
rate $r=const$. To get insight into the role of the chain we consider
first the situation when only one bond is breakable and take this bond
to be either at the fixed or at the pulled end of the chain. The rest
of the chain is considered as a harmonic Rouse one. The influence of
the chain on the breaking properties of the bond is twofold: First,
due to the coupled dynamics, the overall noise force acting on the
monomer stems from the whole rest of the chain and is
non-Markovian. Second, since the force does not propagate immediately
through the chain, a bond at the grafted end of the chain experiences
at the beginning the force smaller then the one that is applied at the
pulled terminal. The rupture of a single breakable Morse bond at the
fixed wall is affected both by the non-Markovian fluctuations and by
the force profile propagation. In contrast, a bond situated at the
pulled terminal of the chain feels the instantaneous force, and the
deviations from the single-bond rupture statistics are solely due to
the non-Markovian character of the noise. Fig.~\ref{fig:mean_n} shows
the most probable rupture force $f_{max}$ as a function of the number
of bonds in the chain $N$ for both situations. The reference value of
the single bond rupture setup is also shown. The difference between
different situations is evident. Thus, for a breakable bond at a wall
$f_{max}$ lies well above the reference value (dotted line), while for
the breakable bond at a pulled terminal $f_{max}$ lies slightly below
the value of the single bond rupture measurement when the breakable
bond is placed at the pulled terminal (dashed line). We conclude that
non-Markovian fluctuations accelerate the rupture process while the
retarded force propagation delays it. Since the effect of delay is by
an order of magnitude stronger, we neglect the impact of the
non-Markovian character of the fluctuations and neglect correlations
introduced by noise by assuming that the rupture of different bonds is
independent.

We pass to a chain of $N$ breakable Morse bonds.  Let $W_1(f_i(t))$ be
the probability of the bond $i$ to be intact. Let $r$ be very small,
$r\ll 1$. Then, we can assume the forces acting on each spring along
the polymer chain are virtually the same. This is expected to be true
at least for a not too large value of $N$. The probability that all
$N$ bonds are intact is then given by $W_{N}(f)=W_1(f)^{N}$.  Then,
the probability that a bond breaks in an interval $[f,f+df]$ is
$P_N(f)=NW_1(f)^{N-1}P_1(f)$ and is given by the same expression as $P_1$ with $v$ changed for $Nv$.  Together
with \eqref{eq:scaling} we derive the following limiting scaling
relation for the most probable rupture force
\begin{equation}
f_{max}=f_i\left[1-\left(\frac{\ln\left(Nv/r\right)}{w}\right)^{\frac{2}{3}}\right]\,.
\label{eq:scaling_rabe_n}
\end{equation}
The experimentally obtained curves of the most probable rupture forces
are expected to tend to this scaling relation in the limit of
vanishing loading rates.

Let us now turn to higher loading rates.  The probability that all $N$
bonds in the chain are still intact under a pulling force at the
terminal of the chain being $f(t)$ is then given by
$W_N(f_i(t))=\prod_{i=1}^N W_1(f_i(t))$.  Passing to the continuum
limit ($i\rightarrow x$) we obtain
\begin{equation}
W_N(t)=\exp\left\{\int_0^N\ln\left(W_1(f(x,t))\right)dx\right\}\,.
\label{eq:wl}
\end{equation}
Since barrier crossing events are very rare, most of the time
the motion of the monomers takes place close to the quadratic potential minima
of bond energies. 
Therefore we derive the force profile by considering a semi-infinite chain of harmonic springs pulled at $x=0$ with a force $f(t)=rt$, i.e., by solving the following
continuum equation for the scalar displacement field $q(x,t)$
\begin{equation}
\dot{q}(x,t)=c\alpha \Delta_x q(x,t)+rt\delta(x)\,.
\label{eq:heat}
\end{equation}
For not too short chains the impact of the clamped end can be neglected. 

Then, the force profile $f(x,t)$ connected to $q(x,t)$ via 
$f(x,t)=-c\alpha {dq}/{dx}$ is given by
\begin{equation}
f(\xi,t)=f(t)\left[
\left(1-\erf\left(\frac{\xi}{2}\right)\right)\left(1+\frac{\xi^2}{2}\right)-
\frac{\xi}{\sqrt{\pi}} \e^{-\frac{\xi^2}{4}}\right]\,,
\label{eq:fl4}
\end{equation}
with $\xi=x/\sqrt{c\alpha t}$. Linearizing this result close to the pulled end,
\begin{equation}
f(\xi,t)\simeq f(t)\left(1-2\xi/\sqrt{\pi}\right)\, ,
\label{eq:fl5}
\end{equation}
and using Eqs. \eqref{eq:w1} and \eqref{eq:wl} we get:
\begin{equation}
\begin{split}
& W_N(f(t))= \exp \left\{-\frac{vl}{3rw^{2/3}\sqrt{f(t)}}\times\right.\\
& \left. \left[\Gamma\left(\frac{2}{3},a(t)\right)-\Gamma\left(\frac{2}{3},a(t)(1+S(N))^{\frac{3}{2}}\right)\right]\right\}\,.
\end{split}
\label{eq:wl3}
\end{equation}
Here $l=f_i\sqrt{c\alpha\pi/r}$, $\Gamma(b,z)$ is the upper incomplete gamma function 
$\Gamma(b,z)=\int_z^\infty t^{b-1}\e^{-t} dt$,
$S(N,f)=2N \sqrt{r/c\alpha \pi f(t)}/[f_i/f(t)-1]$ and $a(t)=w(1-f(t)/f_i)^{3/2}$.
Finally, the probability density distribution reads
\begin{equation}
\label{eq:pdfn1}
\begin{split}
&P_{N}(f(t))=-W_{N}(f(t))\left\{\frac{vl}{6w^{2/3}f(t)^{3/2}}\times\right.\\ 
&\left[\Gamma\left(\frac{2}{3},a(t)\right)-\Gamma\left(\frac{2}{3},a(t)(1+S(N,f))^{\frac{3}{2}}\right)\right]\\ 
&
+\frac{vl}{2\sqrt{f(t)}}\times\\
&\left.\left[\frac{\e^{-a(t)}}{f_{i}}-\left(\frac{1}{f_i}-\frac{N/l}{\sqrt{f(t)}}\right)\e^{-a(t)(1+S(N,f))^{\frac{3}{2}}}\right]\right\}\,.
\end{split}
\end{equation}
\begin{figure}
\begin{center}
\includegraphics[height=6.cm,width=7.cm]{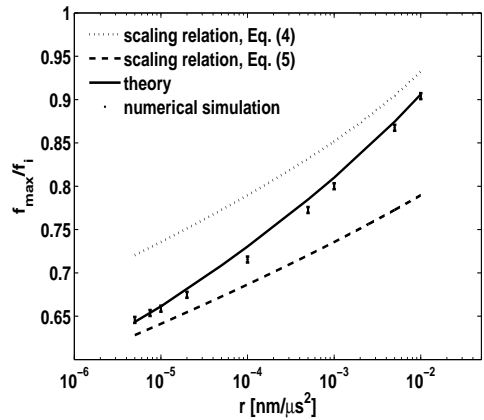}
\caption{\label{fig:scaling_r} Most probable rupture force $f_{max}$ as a function of the loading rate $r$. 
The length of the chain is $N=100$. The remaining parameter values are the same as in Fig.~\ref{fig:mean_n}.}
\end{center}
\end{figure}

In Fig.~\ref{fig:scaling_r} we present the numerically obtained most
probable rupture force $f_{max}$ as a function of the loading rate $r$
for a chain of length $N=100$. Superimposed are the prediction of
Eq.~\eqref{eq:scaling} (dashed line), the prediction of
Eq.~\eqref{eq:scaling_rabe_n} (dotted line), and the most probable
rupture force derived from Eq.~\eqref{eq:pdfn1}. In the limit of small $r$
the most probable rupture force tends to the prediction of
Eq.~\eqref{eq:scaling_rabe_n}: Virtually all bonds account for the
rupture process of the chain. In the opposite limit of very large
loading rates only a few bonds contribute to rupture (in the extreme
limit only the one at the pulled end of the chain) and the scaling  of
$f_{max}$ is given by Eq.~\eqref{eq:scaling}. The crossover behavior 
is very well reproduced by the theory (Eq.~\eqref{eq:pdfn1}). 
Small deviations appear for the intermediate
values of $r$ where the exact force profile along the chain plays a
role and the linear approximation in Eq.~\eqref{eq:fl5} gets
slightly inaccurate. Higher order corrections might resolve this issue.

\begin{figure}
\begin{center}
\includegraphics[height=6.cm,width=7.cm]{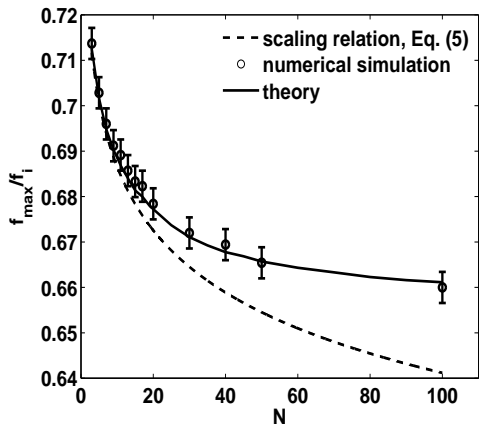}
\end{center}
\caption{\label{fig:rabe_length} Most probable rupture force $f_{max}$ as a function of the chain length $N$ for $r=10^{-5}nm/\mu s^2$. 
The remaining parameter values are the same as in Fig.~\ref{fig:mean_n}.}
\end{figure}

In Fig.~\ref{fig:rabe_length} we present the numerically obtained most
probable rupture force $f_{max}$ as a function of the chain length $N$
for $r=10^{-5}nm/\mu s^2$. Shown are also the prediction of Eq.~\eqref{eq:scaling_rabe_n} (dashed line), and the
most probable rupture force derived from Eq.~\eqref{eq:pdfn1}. For not
too large systems, the predictions of Eq.~\eqref{eq:scaling_rabe_n}
and of Eq.~\eqref{eq:pdfn1} tend to each other and both describe well
the results from the numerical simulations. Moreover, it is evident
that the scaling relation Eq.~\eqref{eq:scaling_rabe_n} gives a lower
bound for the most probable rupture force. Deviations appear for
larger system sizes $N$, where both the numerical results and the
prediction of Eq.~\eqref{eq:pdfn1} saturate. 
We also state that the PDF of the actual rupture
forces (not shown) is also adequately described by
Eq.~\eqref{eq:pdfn1}.

The overall behavior shown in Fig.~\ref{fig:rabe_length} corresponds to a transition from
the small-$N$ scaling regime to a saturation after some $N_s(r)$
depending on the loading rate. The value of $N_s$ can be estimated as
follows. A single bond rupture is governed by $P_1(f)$ and rupture
occurs with highest probability at $f(x,t)=f_{max}$ with $f_{max}$
given in Eq.~\eqref{eq:scaling}. The value of $f_{max}$ can be
translated into the most probable time to break via $t_{max} =
f_{max}/r$.  The distribution $P_1(f)$ is strongly skewed to the left
\cite{Dudko03,Dudko06}, and its variance is given by
$\sigma_1^2=(2\pi^2/27)f_i^2w^{-4/3}\left(\ln\left(v\right)\right)^{-2/3}$,
see Ref.~\cite{Garg95}.  We then can assume that the rupture of a bond
hardly occurs if the corresponding force is $f < f_{max} -
2\sigma_1$. The portion of the chain $N_s$ in which the broken bond is
localized is then determined by the condition that the force at the
bond number $N_s$ is $f_{max} - 2\sigma_1$ at the time when the first
bond is most probably going to break,
i.e. $f(N_s,t_{max})=f_{max}-2\sigma_1$.  resolving Eq.~\eqref{eq:fl5}
for the corresponding value of $x=N_s$ we get
\begin{equation}
N_s(r)=\frac{\sqrt{\frac{c^2\alpha\pi^3}{54r}}}{\left(\ln\left(v\right)\right)^{\frac{1}{3}}w^{\frac{2}{3}}\sqrt{1-\left(\frac{\ln\left(v\right)}{w}\right)^{\frac{2}{3}}}}\,.
\label{eq:ns}
\end{equation}
For $r=10^{-5}nm/\mu s^2$ we have $N_s\simeq 100$, in
qualitative agreement with the outcome of our numerical simulations,
and for $r=10^{-4}nm/\mu s^2$ we get $N_s\simeq 35$ (again in agreement with simulations, not shown).

Let us summarize our findings. Compared to single bond breaking, 
the existence of the chain introduces new aspects into
rupture dynamics, the most important being
the delayed stress propagation along the chain. 
We show that the most probable rupture force decreases with the length of the
chain  as $f_{max}\propto const-(\ln (N/r))^{2/3}$ and then saturates at the value depending on 
the loading rate. 
The results of theoretical considerations are confirmed by the ones
of numerical simulations. 

\begin{acknowledgments}
The authors thankfully acknowledge valuable discussions with J. Klafter and M. Urbakh.
This research has been supported by DFG within the SFB 555 research collaboration program.
\end{acknowledgments}


\end{document}